\documentclass[aps,pra,showpacs,groupedaddress,twocolumn,preprintnumbers,amsmath,amssymb]{revtex4}
\usepackage{graphicx}
\usepackage{dcolumn}
\usepackage{bm}
\usepackage{braket}
\usepackage{color}

\begin{document}

\title{Suppression of relative flow by multiple domains in two-component Bose-Einstein condensates}

\author{Yujiro Eto$^{1}$}
\author{Masaya Kunimi$^{2}$}
\author{Hidekatsu Tokita$^{1}$}
\author{Hiroki Saito$^{2}$}
\author{Takuya Hirano$^{1}$}
\affiliation{%
$^{1}$Department of Physics, Gakushuin University, Toshima, Tokyo 171-8588, Japan\\
$^{2}$Department of Engineering Science, University of Electro-Communications, Chofu, Tokyo 182-8585, Japan}

\date{\today}
             
\begin{abstract}
We investigate flow properties of immiscible Bose-Einstein condensates composed of two different Zeeman spin states of $^{87}$Rb.
Spatially overlapping two condensates in the optical trap are prepared by application of a resonant radio frequency pulse, 
and then the magnetic field gradient is applied in order to produce the atomic flow.
We find that the spontaneous multiple domain formation arising from the immiscible nature drastically changes the fluidity. 
The homogeneously overlapping condensates readily separate under the magnetic field gradient, and they form stable configuration composed of the two layers.
In contrast, the relative flow between two condensates are largely suppressed in the case where the magnetic field gradient is applied after spontaneous domain formation.
\end{abstract}


\maketitle
\section{Introduction} 
Formation of heterogeneous spatial structures in binary phase-separating fluids and their fluidity are intriguing objects of research in the various fields such as non-equilibrium physics, fluid mechanics, and rheology \cite{Onuki97}.
Multi-component atomic Bose-Einstein condensates (BECs) realized in recent decades are quantum fluids that display phase separation \cite{Myatt97}. 
The unique feature of atomic BECs is the unprecedented controllability of relevant experimental parameters, which provides novel approaches for studying the physics of phase separation.
For example, atomic interaction strengths can be tuned via Feshbach resonances \cite{Inouye98, Thalhammer08},
and trapping potentials can be designed by using optical and magnetic fields.
To date, miscibility and immiscibility of binary BECs were controlled by changing the atomic interaction strengths \cite{Papp08, Tojo10} and by using Rabi coupling \cite{Nicklas11,Nicklas14}.
Additionally, quantum tunneling across immiscible BECs separated into two layers was observed by application of a state dependent potential \cite{Kurn99}.
Theoretically pattern formations at fluid interfaces have been predicted in two-component BECs, which are due to interface instabilities such as Rayleigh-Taylor \cite{Sasaki09, Kadokura12} and Kelvin-Helmholtz instabilities \cite{Takeuchi10,Suzuki10}.

When two spatially overlapping immiscible BECs are prepared, domain structures are spontaneously formed so that the free energy of the whole system is minimized \cite{Timmermans98,Trippenbach00,Coen01}.
This spontaneous spatial structure formation has been widely observed in dual species BECs \cite{Papp08} and two-component BECs composed of two different hyperfine states \cite{Myatt97, Hall98, Mertes07} or Zeeman states \cite{Stenger98, Miesner99, Kurn99}.
In this paper, we investigate influences of spontaneous structure formation on the fluidity of immiscible two-component BECs,
and conversely influences of non-equilibrium fluid flow on the domain structures. 
Spatially overlapping two-component BECs comprised of different Zeeman states of $^{87}$Rb atoms are prepared by application of a resonant radio frequency (rf) pulse.
The two-component BECs homogeneously overlap just after the rf pulse, and domain structures grow distinctly after 40 ms.
In order to generate the relative flow between the two-component BECs, a spin-dependent potential gradient is applied.
We find that the center of mass (c.m.) motion is drastically changed depending on whether the two-component BECs are homogeneously overlapped or multiple domain structures are formed.
When two-component BECs are homogeneously overlapped, they readily separate into two layers, which is a stable configuration for immiscible fluids. 
In contrast, multiple domains refuse to separate against the spin-dependent potential gradient, and the relative flow is suppressed.

In the previous experiments in Refs. \cite{Maddaloni00, Modugno00}, the collective dynamics of immiscible BECs in a highly displaced magnetic harmonic trap have been observed.
The effects of periodic collisions between two-component BECs on c.m. oscillations and shape oscillations were investigated.
On the other hand, our objective is to reveal the influence of spontaneous domain formation on the collective motion by controlling the timing of application of a spin-dependent gradient potential.

The paper is organized as follows. In Sec. II, we describe our experimental and theoretical methods. 
The results and discussions are provided in Sec. III, and conclusions are given in Sec. IV.  

\section{Experimental and Theoretical Procedures}
A BEC of $4 \times 10^5$ $^{87}$Rb atoms in the hyperfine state $ \ket{F = 2, m_{F} = -2} $ is produced  in a crossed far off-resonant optical dipole trap (FORT) with axial ($z$-direction) and radial frequencies of $\omega_z / (2\pi) = 32$ Hz and $\omega_{\rm r} / (2\pi) = 135$ Hz.
A more detailed description of creation of the BEC is given in Ref. \cite{Eto13APEX}.
In order to create the stable bias field $B_{z}$ along the $z$ direction, the whole apparatus to create the BECs is installed inside a magnetic shield room, and a laser diode source with low ripple noise of less than $20$ $\mu$A (Newport, LDX-3232-100V) is used as the current source for the $z$-axis Helmholtz coil.
This ripple noise current is 50 dB less than the current used to create $B_z$.

Two-component BECs composed of $\ket{F = 2, m_{F} = -1} \equiv \ket{1}$ and $ \ket{F = 2, m_{F} = -2} \equiv \ket{2}$ are prepared by application of a resonant rf pulse.
The intra- and interspecies $s$-wave scattering lengths of the components $\ket{1}$ and $\ket{2}$ are $(a_{11}, a_{22}, a_{12}) = (95.68, 98.98, 98.98)a_{\rm B}$ in units of the Bohr radius \cite{Widera06}, which satisfy the immiscible condition, $\sqrt{a_{11} a_{22}} < a_{12}$.
After a holding time of $T_{\mathrm{hold}} =$ $0$ ms or $40$ ms,
the magnetic field gradient, $dB_{z}/dz$, is applied, which creates the $m_F$-dependent potential gradient along the $z$ direction. 
The two-component BECs move under the $m_{F}$-dependent gradient potential during $T_{\mathrm{grad}}$ and then they are released from the FORT.
The two component BECs are then spatially separated along the $z$ direction by the Stern-Gerlach (SG) method.
After a time-of-flight (TOF) of 15 ms, the atomic distribution of each component is measured using absorption imaging.


In the numerical simulations, we use the mean-field approximation. 
The dynamics of the macroscopic wave functions $\Psi_1(\bm{r}, t)$ and $\Psi_2(\bm{r}, t)$, which correspond to the components $\ket{1}$ and $\ket{2}$, are described by the two-component Gross-Pitaevskii (GP) equation:
\begin{align}
i\hbar\frac{\partial}{\partial t}\Psi_j(\bm{r}, t)=\left[-\frac{\hbar^2}{2m}\nabla^2+U_j(\bm{r}, t)+g_{j,j}|\Psi_j(\bm{r}, t)|^2\right.\notag \\
+g_{j,3-j}|\Psi_{3-j}(\bm{r}, t)|^2\biggr]\Psi_j(\bm{r}, t),\quad (j=1,2)\label{eq:two-component_GP}
\end{align}
where $m$ is the mass of a ${}^{87}$Rb atom, 
and $g_{j,k}=g_{k,j}$ is the coupling constant between $\ket{i}$ and $\ket{k}$.  
The harmonic potential and the magnetic field gradient are represented by
$U_j(\bm{r}, t)=m[\omega_{\rm r}^2(x^2+y^2)+\omega_{z}^2z^2]/2-j\mu_{\rm B}(dB_z/dz)\theta(t-T_{\rm hold})z/2$, where $\theta(\cdot)$ is the Heaviside step function and $\mu_{\rm B}$ is the Bohr magneton. 
In order to include the effects of the spin-dependent two-body inelastic loss for $m_F=-1$ component, we use $g_{1,1}=4\pi\hbar^2a_{11}/m-3i\hbar b_2/14$ and $g_{12}=g_{22}=4\pi\hbar^2a_{22}/m$, where $b_2=24.3\times 10^{-14}{\rm cm}^3/{\rm s}$ is the inelastic collision coefficient reported in Ref.~\cite{Tojo2009}.

We solve the two-component GP equation using the pseudospectral method \cite{Pseudospectral}. The initial conditions are prepared as follows: we first obtain the ground state $\Psi(\bm{r})$ of the GP equation (\ref{eq:two-component_GP}) under the condition $\Psi_1(\bm{r}, t)=0$ by the imaginary-time evolution method, where $\Psi(\bm{r})$ is normalized by the number of atoms, $\int d\bm{r}|\Psi(\bm{r})|^2=4\times 10^5$. The real-time GP equation is then solved for the initial condition $\Psi_1(\bm{r}, t=0)=\sqrt{2/3}\Psi(\bm{r})$ and $\Psi_2(\bm{r}, t=0)=\sqrt{1/3}\Psi(\bm{r})$. In the TOF expansion, the harmonic potential in $U_j(\bm{r}, t)$ is switched off. The numerical meshes are typically $64\times 64\times 512$ for in-trap simulations and $512\times 512\times 1024$ for the TOF simulations.


\section{Experimental and Theoretical Results}
\subsection{rf spectroscopy and state preparation}

\begin{figure}
\includegraphics[width=8cm]{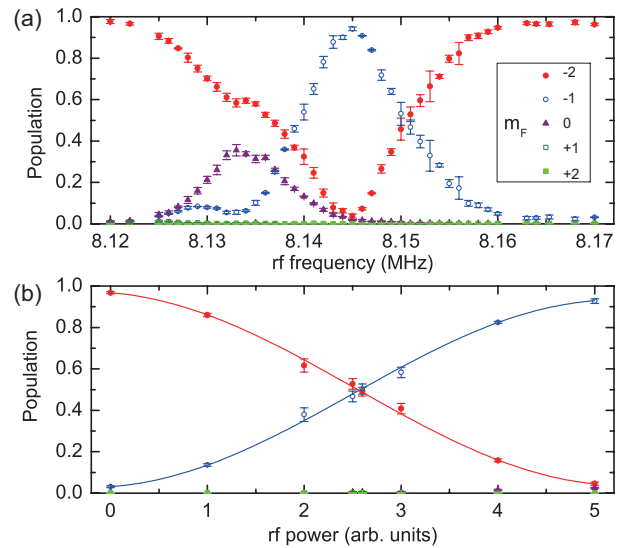}
\caption{(Color online) rf spectroscopy.
The initial state $\ket{2}$ is irradiated by an rf pulse.
The rf pulse shape is a Gaussian function with a standard deviation of $9.5$ $\mu$s.
(a) and (b) show the frequency and intensity dependence of population of each $m_{F}$ component, respectively.
In (b), the frequency of rf pulse is set to be 8.145 MHz.
Each point represents the average over three measurements with the error bars giving the standard deviation over those measurements.}
\label{fig1}
\end{figure}

In order to experimentally prepare two-component BECs comprised of $\ket{1}$ and $\ket{2}$, we carried out rf spectroscopy.
Figure 1 shows the population of each $m_{F}$ component obtained by the application of a single rf pulse with a constant frequency to the initial state $\ket{2}$.
From the results of frequency dependence [Fig. 1(a)],
the resonant frequency of the rf transition between $\ket{1}$ and $\ket{2}$ is found to be $8.145$ MHz.
The two-photon transition into the $m_{F} = 0$ component was also observed around $8.133$ MHz.
Atoms in $m_F = 0$, $+1$ and $+2$ components are not populated for the resonance frequency between $\ket{1}$ and $\ket{2}$ due to the large second-order Zeeman shift and the stable magnetic field environment.
As shown in Fig. 1(b), we can prepare the superposition state of $\ket{1}$ and $\ket{2}$ with an arbitrary population by changing the intensity of rf pulse at $8.145$ MHz.
From these spectroscopic measurements,
the strength of the $B_{z}$ field is found to be $B_{z} = 11.685$ G.

\subsection{Spontaneous multiple domain formation in an immiscible two-component BECs}

\begin{figure}
\includegraphics[width=8cm]{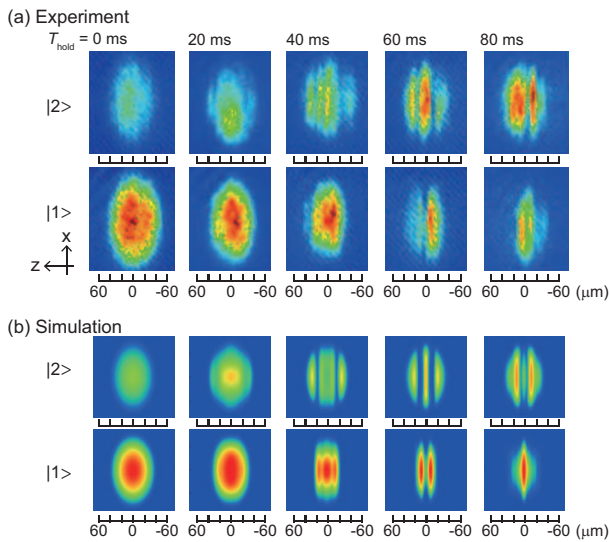}
\caption{(Color online) Experimentally observed (a) and numerically simulated (b) phase-separation dynamics.
Absorption images are obtained at $T_{\mathrm{hold}} =$ $0$ ms, $20$ ms, $40$ ms, $60$ ms, and $80$ ms from left to right.
The two-component BECs of $\ket{1}$ and $\ket{2}$ states are initially prepared in the ratio of 2:1.
The magnetic field gradient to create the state-dependent potential is not applied in this experiment and numerical simulation.
}
\label{fig2}
\end{figure}

We observed the spontaneous spatial structure formation of two-component BECs composed of $\ket{1}$ and $\ket{2}$.
Figure 2(a) shows the typical absorption images observed for $T_{\mathrm{hold}} =$ $0$ ms, $20$ ms, $40$ ms, $60$ ms, and $80$ ms,
where the initial ratio of the components $\ket{1}$ and $\ket{2}$ is 2:1.
Just after the application of the rf pulse ($T_{\mathrm{hold}} = 0$ ms), the two-component BECs overlap each other [leftmost in Fig. 2(a)]. 
After $T_{\mathrm{hold}}$ = 40 ms, spatial domain structure grows distinctly [middle and right in Fig. 2(a)].

The numerical simulation of the coupled GP equation is shown in Fig. 2(b).
Though the growth of multiple domain structure is reasonably reproduced in the numerical simulation,
the experimentally observed patterns are asymmetric along the z-axis.
A possible cause is the asymmetry of the trapping potential due to experimental imperfections such as small optical misalignments.
In addition, the experimentally observed spatial structures are not clear due to the finite spatial resolution of our imaging system ($\sim 7.5$ $\mu$m).

\begin{figure}
\includegraphics[width=8cm]{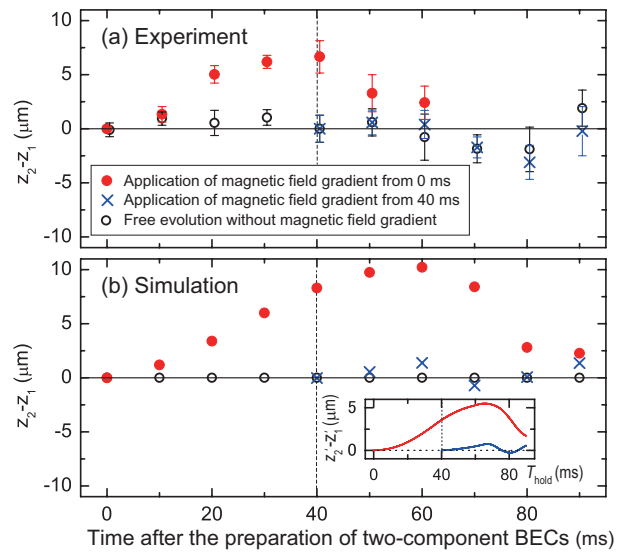}
\caption{(Color online) Relative c.m. motion between the two-component BECs, $z_2-z_1$, as a function of time after the preparation of two-component BECs.
The open circles represent the data calculated from results of Fig. 2, namely they correspond to the free evolution without magnetic field gradient.
The filled circles (cross marks) are measured by changing $T_{\mathrm{grad}}$ with the fixed value of $T_{\mathrm{hold}} = 0$ ms ($T_{\mathrm{hold}}= 40$ ms), where the magnetic field gradient of $15$ mG/cm is applied during $T_{\mathrm{grad}}$. 
The c.m. positions after TOF differ from those in the trap, and their values might be affected by the in-trap velocity.
However the red points in (b) seem to scale those of in-trap [Inset in (b)] thanks to the short TOF time of $15$ ms and small magnetic field gradient of $15$ mG/cm.
}
\label{fig3}
\end{figure}

\subsection{Center of mass motion under the state dependent potential gradient}

\begin{figure*}
\includegraphics[width=16cm]{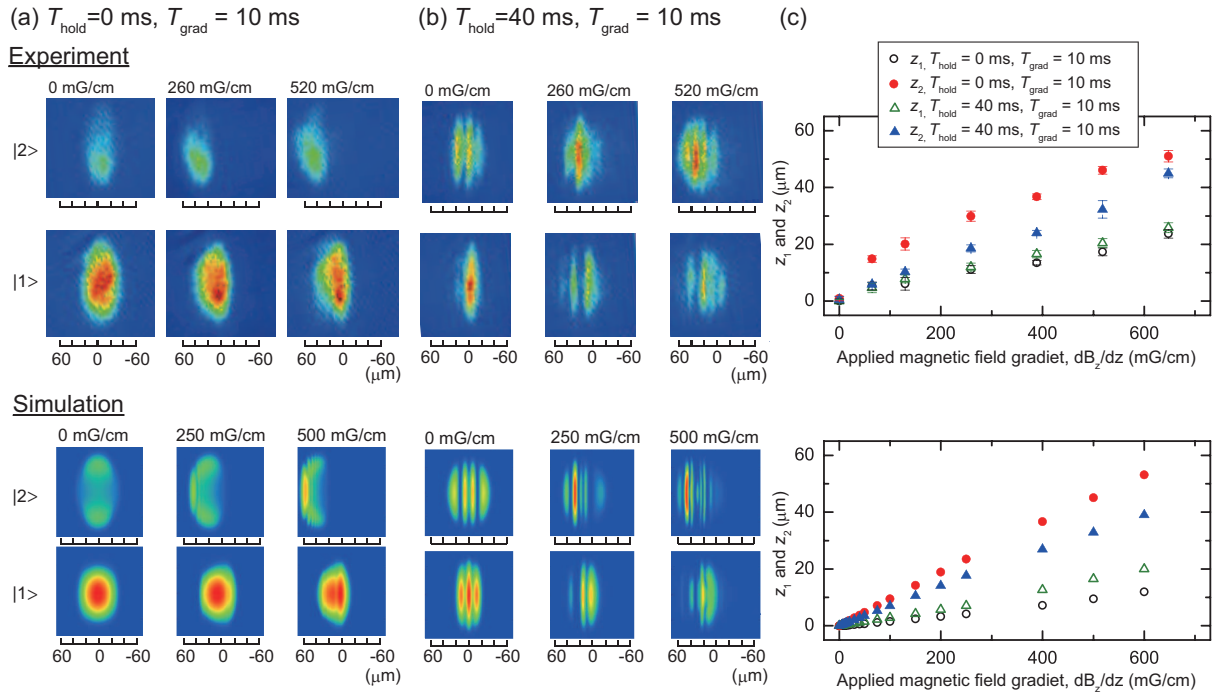}
\caption{(Color online) Structural change and c.m. motion induced by the application of magnetic field gradient, 
where the distance traveled by SG measurement is compensated. 
(a) Typical absorption images obtained at $T_{\mathrm{hold}} = 0$ ms and $T_{\mathrm{grad}} = 10$ ms.
(b) Typical absorption images obtained at $T_{\mathrm{hold}} = 40$ ms and $T_{\mathrm{grad}} = 10$ ms.
(c) c.m. positions ($z_1$ and $z_2$) calculated from data of (a) and (b) versus the applied magnetic field gradient.
Top and bottom panels represent the experimental and numerical results, respectively.
In the experimental data, the magnitude of the effective magnetic field gradient is larger than the horizontal axis in (c) due to the overshoot of the applied magnetic field gradient.
See Supplemental Material for movies showing time evolution of the two-component BECs in a trap potential \cite{suppl}.}
\label{fig4}
\end{figure*}

Figure 3 shows the experimental and numerical results of the relative c.m. motion between two-component BECs, $z_{2} - z_{1}$, with and without application of the magnetic field gradient.
In order to investigate the effect of the $m_F$-dependent potential gradient $\propto -m_{F}dB_{z}/dz$ on two-component BECs without the domain structure,
we measure the dynamics in two different cases: (i) the free evolution without magnetic field gradient (the open circles) and (ii) application of magnetic field gradient of $15$ mG/cm just after the preparation of two-component BECs (the filled circles).
We define that $T_{\mathrm{grad}}$ and $T_{\mathrm{hold}}$ are the time periods with and without the magnetic field gradient, respectively. 
Thus, $T_{\mathrm{hold}}$ is varied with $T_{\mathrm{grad}} = 0$ ms in (i), and $T_{\mathrm{grad}}$ is varied with $T_{\mathrm{hold}} = 0$ ms in (ii).
In both of experiments and simulations, the relative c.m. of the two-component BECs is almost unchanged for the case (i), 
whereas the c.m. of the two-component BECs is clearly separated for the case (ii). 
In (ii), although the field gradient pushes the two-component BECs in the same direction, component $\ket{1}$ moves in the direction opposite to the gradient force.
This is because component $\ket{1}$ is pushed away by component $\ket{2}$ due to the repulsive interaction between the two components \cite{Maddaloni00, Modugno00}.
After $T_{\mathrm{hold}} \simeq 40$ ms, the relative c.m. decreases due to the harmonic potential. 
This period is longer than the free oscillation period in the trap due to the inter-component interaction.

In order to investigate the effect of the domain structure on the c.m. motion, we observed the dynamics for the case (iii): after $T_{\mathrm{hold}} = 40$ ms the magnetic field gradient of $15$ mG/cm is applied during $T_{\mathrm{grad}}$ 
(the cross marks in Fig. 3 are obtained by varying $T_{\mathrm{grad}}$).
Since the spatial domain structure is observed clearly at $T_{\mathrm{hold}} =$ $40$ ms as shown in Fig. 2(a), the magnetic field gradient is applied after the domain structure is established in the case (iii).
As well as the case (i), the relative c.m. positions are almost unchanged for the case (iii).
This result indicates that the spontaneous domain formation suppresses the relative motion between two-component BECs.
The quantitative differences between experiments and simulations in Fig. 3 are likely due to the asymmetric pattern formation observed in Fig. 2, 
because the quantitative differences largely occur after $40$ ms.

Figures 4(a) and 4(b) show the typical absorption images obtained for various magnitudes of applied magnetic field gradient.
When $T_{\mathrm{hold}} = 0$ ms and $T_{\mathrm{grad}} = 10$ ms [Fig. 4(a)], in which the magnetic field gradient is applied just after the preparation of the two-component BECs,
two-component BECs are readily separated.
The resulting spatial structures consist of two layers, which is a stable configuration for the immiscible fluids under a spin dependent gradient potential.
In the case of $T_{\mathrm{hold}} = 40$ ms and $T_{\mathrm{grad}} = 10$ ms [Fig. 4(b)], even for a large gradient about $500$ mG/cm, 
the two-component BECs move while maintaining multiple domain structure so that the spatial overlap between the components $\ket{1}$ and $\ket{2}$ is not increased.  

In order to quantitatively evaluate these motions, $z_1$ and $z_{2}$ are calculated from Figs. 4(a) and 4(b), which are shown in  Fig. 4(c). 
Compared with case of $T_{\mathrm{hold}} = 0$ ms (the open and filled circles),
$z_1$ moves faster and $z_2$ moves slower in the case of $T_{\mathrm{hold}} = 40$ ms (the open and filled triangles).
This result indicates that the multiple domain structure suppresses the separation of the two-component BECs, even when the magnetic field gradient is $500$ mG/cm.
In other words, one of the two-component BECs serves as the potential barrier for the other BEC in the presence of the multiple domains.
However, the domain structure does not completely suppress the change in $z_1 - z_2$.
According to the simulations \cite{suppl}, for our experimental parameters, this is because the two-component BECs pass by each other, rather than the quantum tunneling \cite{Kurn99}.
In the case that the magnetic field gradient of $500$ mG/cm is applied from $T_{\mathrm{hold}} = 40$ ms, after $T_{\mathrm{grad}} \sim 60$ ms, 
two-component BECs are spatially separated into two layers and the domain structure disappears.

Finally we discuss the fluidity under the larger magnetic field gradient.
When the typical interaction energy $g_{12}(n_1+n_2)$ and Zeeman energy $\mu_B(dB_z/d z)w_{\rm domain}/2$ are comparable, 
the effects of the domain become less important, where $n_1$ and $n_2$ are the particle densities of the components $\ket{1}$ and $\ket{2}$, respectively, and $w_{\rm domain}$ is a typical domain size $10 \sim 20$ $\mu{\rm m}$. 
According to this estimation, we find that the effects of the domain becomes irrelevant when $dB_z/d z\sim 2000{\rm mG/cm}$.

\section{Conclusions}
We have investigated the flow properties of immiscible two-component BECs composed of two different Zeeman states of $^{87}$Rb.
The two immiscible BECs evolve in the optical trap forming multiple domain structures.
In order to investigate the effect of these multiple domains on the fluidity,
 a spin dependent gradient potential, which produces an atomic flow, was applied at different times.  
We find that the induced flow is significantly affected by the spatial structures. 
At early times, the spatially overlapping condensates readily separate under the state dependent potential gradient. 
However, after sufficient time is given to allow domain formation, the two-component BECs are difficult to separate to each other, since the multiple domains play role of barriers that prevent the BECs to separate. 
Our results demonstrate the complex nature of flow in two-component immiscible BECs. 
Despite the intuitive expectation that immiscible components should separate under a spin dependent gradient potential, 
multiple domains arising from the immiscible nature actually resist the relative flow of the two-component BECs.
We anticipate that the method we have used here to probe the effects of domain formation can be applied to further studies of quantum fluids in the future.

\begin{acknowledgments}
We would like to thank M. Sadgrove for his valuable comments.
This work was supported by MEXT KAKENHI Grant Number 25103007 and JSPS KAKENHI Grant Number 26400414 and 15K05233.
\end{acknowledgments}

\end{document}